\documentclass[letterpaper,11pt]{article}
\pdfoutput=1

\usepackage{jheppub}
\usepackage{braket}
\usepackage{empheq}

\renewcommand{\b}[1]{\textbf{#1}}

\newcommand{\al}{\alpha}
\newcommand{\an}[1]{\left\langle#1\right\rangle}
\newcommand\D{\Delta}
\newcommand{\dhat}{\Hat{\mathcal{D}}}
\newcommand\e{\epsilon}
\newcommand{\mA}{\mathcal{A}}
\newcommand\mO{\mathcal{O}}
\newcommand{\rmd}{\mathrm{d}}
\newcommand{\sq}[1]{\left[#1\right]}
\newcommand{\V}{\mathcal{V}}
\newcommand{\zb}{\bar{z}}

\title{Loop-level gluon OPEs in celestial holography}

\author[a]{Rishabh Bhardwaj,}
\author[a]{Luke Lippstreu,}
\author[a]{Lecheng Ren,}
\author[a,b]{Marcus Spradlin,}
\author[a]{Akshay Yelleshpur Srikant}
\author[a]{and Anastasia Volovich}

\affiliation[a]{Department of Physics,
	Brown University,
	Providence, RI 02912, USA}
\affiliation[b]{Brown Theoretical Physics Center,
	Brown University,
	Providence, RI 02912, USA}

\emailAdd{rishabh\_bhardwaj@brown.edu}
\emailAdd{luke\_lippstreu@brown.edu}
\emailAdd{lecheng\_ren@brown.edu}
\emailAdd{marcus\_spradlin@brown.edu}
\emailAdd{akshay\_yelleshpur\_srikant@brown.edu}
\emailAdd{anastasia\_volovich@brown.edu}

\abstract{We compute one-loop corrections to the OPE of gluons in the celestial conformal field theory corresponding to Yang-Mills coupled to arbitrary matter. We exploit universal hard/soft factorization to derive an IR finite OPE for the hard gluon operators. This OPE involves logarithms and operators that resemble logarithmic partners of primary operators. We derive an exact all-loop OPE in a limit of the Higgs-regulated planar $\mathcal{N}=4$ super Yang-Mills theory.}

\begin{document}

\maketitle

\section{Introduction}

The celestial holography framework~\cite{Pasterski:2016qvg,Pasterski:2017kqt,Pasterski:2017ylz} (see~\cite{Raclariu:2021zjz,Pasterski:2021rjz,Pasterski:2021raf,McLoughlin:2022ljp} for recent reviews) suggests interpreting four-dimensional scattering amplitudes as correlation functions (also called ``celestial amplitudes'') of primary operators in a celestial conformal field theory (CCFT). It seems clear that any putative CCFT would have some quite mysterious features, but this framework has revealed interesting previously hidden features of scattering amplitudes, such as the fact that the infinite tower of tree-level soft graviton symmetries generate a symmetry algebra with $w_{1 + \infty}$ structure, under which soft gluon symmetries also transform irreducibly~\cite{Guevara:2021abz,Strominger:2021lvk}.

A central tool in the study of CFT is the operator product expansion (OPE); in particular the $w_{1 + \infty}$ structure is imprinted in celestial OPE coefficients. Celestial amplitudes for gluons at tree level were considered first in~\cite{Pasterski:2017ylz}, and the OPE of celestial gluon primaries was worked out in~\cite{Fan:2019emx,Pate:2019lpp}. Aspects of loop corrections to gluon amplitudes have been studied in several papers in~\cite{Banerjee:2017jeg,Albayrak:2020saa,Gonzalez:2020tpi,Gonzalez:2021dxw,Nastase:2021izh}. In particular, in~\cite{Gonzalez:2020tpi} it was noted that in dimensionally-regulated planar $\mathcal{N}=4$ super-Yang-Mills (SYM) theory, loop corrected celestial gluon amplitudes can be expressed as differential operators acting on the corresponding tree-level celestial amplitude.

This paper has two parts.  In the first part we use one-loop corrections to splitting functions to compute one-loop corrections to the OPE of two positive helicity gluons in dimensionally regulated Yang-Mills theory coupled to arbitrary (fundamental) matter.  The result, shown in~(\ref{eq:++OPEfinal}), has a piece that can be expressed as a differential operator acting on the tree-level OPE, but for general matter content also has a second term involving the negative helicity gluon. Using the hard/soft factorization theorem for massless gauge theories, whose implication for celestial amplitudes was analyzed in~\cite{Gonzalez:2021dxw}, we can subtract the universal IR divergent terms. The resulting finite OPE of the hard operators is shown in~(\ref{eq:hardOPEfinal}).

In the second part we study the amplitudes of massless gluons in a particular corner of the Coulomb branch of planar $\mathcal{N}=4$ super-Yang-Mills (SYM) theory that has received attention for example in~\cite{Alday:2009zm,Henn:2010bk,Henn:2010ir}. The motivation is two-fold.  First, these Higgs-regulated amplitudes are infrared finite, with a physical mass parameter $m$ replacing the dimensional regularization parameter $\epsilon$ whose interpretation in the context of celestial holography is rather unclear. Second, the four- and five-gluon amplitudes are believed to be known exactly (to all loop order) in the limit when $m^2$ is very small compared to the Mandelstam invariants---they are given by exponential formulas reminiscent of the BDS ansatz~\cite{Bern:2005iz} in dimensional regularization, whose celestial incarnation was computed in~\cite{Gonzalez:2020tpi}. Indeed, the computation of the Higgs-regulated celestial four-gluon amplitude is essentially identical to that of~\cite{Gonzalez:2020tpi}; see appendix~A. We derive the all-loop OPE for two positive helicity gluon operators in the Higgs-regulated theory in~(\ref{eq:higgsalllloopope}).

One reason to be interested in loop-corrected or exact OPEs is to see whether and how they exhibit $w_{1+\infty}$ structure. In the case of self-dual gravity, for example, it is known that the $w_{1+\infty}$ symmetry is perturbatively exact~\cite{Ball:2021tmb}. Unfortunately the loop-corrected gluon OPEs computed in this paper contain logarithms that obstruct the usual CFT mode expansion and therefore complicate the analysis of the symmetry algebra; this remains an interesting problem for future work.

This paper is organized as follows. In section~\ref{sec:collinearlimits} we briefly review the well-known one-loop splitting functions in gauge theory with various matter content. In section~\ref{sec:OPEs} we use these splitting functions to compute one-loop corrections to the OPE of celestial gluon operators and utilize the hard/soft factorization to define a finite OPE for hard gluon operators. In section~\ref{sec:allloopOPEs} we compute the all-loop OPE in the Higgs-regulated planar $\mathcal{N}=4$ SYM theory. We find that the OPE of hard operators computed in dimensional regularization agrees precisely with the OPE in the Higgs-regulated theory. Finally in section~\ref{sec:discussion} we make some preliminary comments and speculation regarding the potential need for using the machinery of logarithmic CFT for properly understanding these celestial OPEs.

\section{Collinear singularities of one-loop amplitudes in gauge theory}
\label{sec:collinearlimits}

Scattering amplitudes in Yang-Mills theory with gauge group $SU(N_c)$ have a perturbative expansion
\begin{align}
    \label{eq:loopexpansion}
    \mathcal{A}_n^{a_1 \dots a_n} \left(p_1,\sigma_1; \dots p_n,\sigma_n\right) = g^{n-2} \sum_{L=0}^{\infty} \al^L \mathcal{A}_n^{(L)a_1 \dots a_n }\left(p_1,\sigma_1; \dots p_n,\sigma_n\right),
\end{align}
where $g$ is the Yang-Mills coupling constant, $\al = a \left(4\pi e^{-\gamma_E}\right)^{\e}$ with $a= \frac{g^2 N_c}{8\pi^2}$ the 't Hooft coupling, $\epsilon$ the dimensional regularization parameter, $\left\lbrace p_1, \dots p_n \right\rbrace$ the momenta of the external gluons, $\sigma_i = \pm 1$ their helicities, and $a_i$ the color indices. They develop singularities when two particles become collinear. If $p_1, p_2$ are becoming collinear, we can parameterize them\footnote{We will only consider the collinear limit of two outgoing gluons.} by
\begin{align}
    \label{eq:collinearlimitpar1}
    p_1 = t \,P\,, \qquad\qquad p_2 = \left(1-t\right)P\,.
\end{align}
In this limit the tree-level amplitude behaves as
\begin{multline}
    \label{eq:collineartree}
    \mA_n^{(0)a_1a_2 \dots a_n} \left(p_1,\sigma_1;p_2,\sigma_2; \dots p_n,\sigma_n\right) \\
    \overset{1\parallel2}{\longrightarrow} \, if^{a_1a_2a_p}\, \sum_{\sigma = \pm }\,\text{Split}^{(0)}_{-\sigma}(t;1^{\sigma_1},2^{\sigma_{2}})\,\mA_{n-1}^{(0)a_p a_3\dots a_n}(P,\sigma;p_3,\sigma_3;\dots)\,.
\end{multline}
The singular behavior is entirely captured by the splitting functions~\cite{Mangano:1987xk,Berends:1987me}
\begin{alignat}{2}
    &\text{Split}^{(0)}_{-}(t;1^{-},2^{-})=0\,,\qquad &&\text{Split}^{(0)}_{-}(t;1^{+},2^{+})=\frac{1}{\sqrt{t(1-t)}\braket{12}}\,,\\
    &\text{Split}^{(0)}_{-}(t;1^{+},2^{-})=\frac{t^2}{\sqrt{t(1-t)}[12]}\,,\qquad &&\text{Split}^{(0)}_{-}(t;1^{-},2^{+})=\frac{(1-t)^2}{\sqrt{t(1-t)}[12]}\label{treesplit}\,,
\end{alignat}
where $\an{12}$ and $\sq{12}$ are the usual spinor brackets. Loop amplitudes exhibit a similar behavior~\cite{Bern:1994zx,Bern:1996je,Bern:1995ix,Mangano:1990by,Dixon:1996wi,Bern:1999ry,Kosower:1999rx,Kosower:1999xi,Bern:2004cz}, captured by
\begin{multline}
\label{eq:collinearloop}
   \mathcal{A}^{(1)a_1a_2a_3...a_n}_n\overset{1\parallel 2}{\longrightarrow}if^{a_1a_2a_p}\sum_{\sigma = \pm }\Big[\text{Split}^{(0)}_{-\sigma}(t;1^{\sigma_1},2^{\sigma_2})\mathcal{A}_{n-1}^{(1)a_Pa_3...a_n}\\
     + \, \text{Split}^{(1)}_{-\sigma}(t;1^{\sigma_1},2^{\sigma_2})\mathcal{A}_{n-1}^{(0)a_Pa_3...a_n}\Big].
\end{multline}
Here and in what follows we will suppress the momentum and spin dependence of amplitudes where possible. The one-loop splitting functions used in this paper are~\cite{Bern:1994zx,Bern:1994fz,Bern:1995ix}
\begin{align}
    \text{Split}^{(1)}_{-}(t;a^{+},b^{+})&=(G^f+G^n)\text{Split}^{(0)}_{-}(t;a^{+},b^{+})\label{plusplusgoestoplus}\,,\\
      \text{Split}^{(1)}_{-}(t;a^{\pm},b^{\mp})&=G^n\,\text{Split}^{(0)}_{-}(t;a^{\pm},b^{\mp})\,,\\
        \text{Split}^{(1)}_{+}(t;a^{+},b^{+})&=G^f\frac{\sq{ab}}{\sqrt{t(1-t)}\an{ab}^2}\,,\label{effFcubed}
        \end{align}
where the factorizing $G^f$ and non-factorizing $G^{n}$ parts read\footnote{It is sufficient to retain terms up to order $\e^{0}$ in the splitting functions at one loop. We will need to retain terms to higher order in $\e$ while performing higher loop computations. All-order expressions for the one-loop splitting functions can be found in~\cite{Bern:2005iz}.}${}^{,}$\footnote{For outgoing particles $s_{ab}>0$ and the prescription for analytic continuation of the logarithm to negative values is $\ln(-|s_{ab}|)=\ln|s_{ab}|-i\pi$~\cite{Bern:2004cz}.}
\begin{align}
    G^f&=\frac{1}{6}\Big(1+\frac{n_s}{N_c}-\frac{n_f}{N_c}\Big)t(1-t)+\mO(\e)\,,\label{factorizingpiece}\\
    G^n&=\hat{c}_{\Gamma}\Bigg[-\frac{1}{\e^2}\Bigg(\frac{\mu^2}{-s_{ab}t(1-t))}\Bigg)^{\epsilon}+2\ln(t)\ln(1-t)-\frac{\pi^2}{6}\Bigg]+\mathcal{O}(\e)\,,
\end{align}
where $n_s$ and $n_f$ are the numbers of massless scalars and fermions, respectively, presumed to be in the fundamental representation, and
\begin{equation}
\label{eq:chat}
    \hat{c}_{\Gamma} = \frac{e^{\epsilon\gamma}}{2}\frac{\Gamma(1+\e)\Gamma^2(1-\e)}{\Gamma(1-2\e)}\,.
\end{equation}
A complete listing of one-loop splitting functions, including ones for fermions, can be found in~\cite{Bern:1994zx,Bern:1994fz}.

\section{Celestial one-loop OPEs from collinear limits}
\label{sec:OPEs}

Now consider the celestial amplitude $\tilde{\mathcal{A}}_n$ (involving $n$ gluons), which is related to the usual scattering amplitude via a Mellin transform~\cite{Pasterski:2016qvg}
\begin{equation}
    \tilde{\mathcal{A}}_n^{a_1\dots a_n} \left(\D_1,\sigma_1;\dots; \D_n,\sigma_n\right) =\int_0^{\infty}\prod_{i=1}^n\,d\omega_i\,\, \omega_i^{\Delta_{i}-1}\,\mathcal{A}_n^{a_1\dots a_n}\left(p_1,\sigma_1; \dots; p_n,\sigma_n\right),
\end{equation}
where we have parameterized the massless four-momenta of the external gluons as
\begin{equation}
    p_i^{\mu}=\epsilon_i\frac{\Lambda \, \omega}{2}\left(1+z_i \zb_i,z_i+\bar{z}_i,-i(z_i-\bar{z}_i),1-z_i\bar{z}_i\right)\label{eq:momentaparameterization}
\end{equation}
where $\e=-/+$ for incoming/outgoing particles respectively and $\Lambda$ is a new parameter that has dimensions of energy and $\omega$ is kept dimensionless for the Mellin integral to make sense dimensionally. We work in a $(+,-,-,-)$ signature bulk spacetime throughout, and assume all particles to be outgoing. A choice of spinor helicity variables corresponding to this parameterization, and the resulting spinor brackets, are
\begin{align}
\label{eq:anglesquare}
    &\lambda_i = \sqrt{\Lambda\omega_i}\begin{pmatrix} 1 \\ z_i\end{pmatrix}, \qquad \tilde{\lambda}_i = \sqrt{\Lambda\omega_i}\epsilon_i\begin{pmatrix} 1 \\ \zb_i\end{pmatrix}, \\
    &\an{ij} = \Lambda\sqrt{\omega_i \, \omega_j} z_{ij}\,, \qquad \sq{ij} = \Lambda\epsilon_i \epsilon_j \sqrt{\omega_i \, \omega_j}\zb_{ij}\,,\nonumber
\end{align}
where $z_{ij} = z_i - z_j$. It has been conjectured~\cite{Pasterski:2016qvg} that celestial amplitudes can be interpreted as correlation functions in a 2D CFT living on the celestial sphere at null infinity, referred to as celestial conformal field theory (CCFT); choosing a convenient normalization, we have\footnote{Henceforth we omit color indices, where possible, in order to avoid notational clutter.}
\begin{align}
    \label{eq:CCFT}
    \tilde{\mathcal{A}}_n = g^{n-2}\an{\mO_{\D_1,\sigma_1}\left(z_1,\zb_1\right) \dots \mO_{\D_n,\sigma_n}\left(z_n,\zb_n\right)}.
\end{align}
Carrying over the decomposition in~(\ref{eq:loopexpansion}) to celestial amplitudes gives
\begin{align}
    \label{eq:loopexpcelestial}
    &\tilde{\mathcal{A}}_n \left(\D_1,\sigma_1;\dots ;\D_n,\sigma_n\right) = g^{n-2} \sum_{L=0}^{\infty}\alpha^L \tilde{\mA}_n^{(L)}\left(\D_1,\sigma_1;\dots ;\D_n,\sigma_n\right)
\end{align}
where
\begin{align}
    \label{eq:loopcelestialdef}
    \tilde{\mA}_n^{(L)}\left(\D_1,\sigma_1;\dots ;\D_n,\sigma_n\right) = \int \prod_{i=1}^n d\omega_i \, \omega_i^{\D_i-1} \mA_n^{(L)} \left(p_1,\sigma_1; \dots; p_n,\sigma_n\right)
\end{align}
and we identify
\begin{align}
\an{\mO_{\D_1,\sigma_1}\left(z_1,\zb_1\right) \dots \mO_{\D_n,\sigma_n}\left(z_n,\zb_n\right)}^{(L)} := \tilde{\mA}_n^{(L)}\,.
\end{align}

\subsection{OPEs from splitting functions}
\label{subsec:OPEsfromsplitting}

One consequence of the dictionary in~(\ref{eq:CCFT}) is that the OPE is determined by collinear limits of amplitudes. Here we work out the general form of this relation. In a nutshell, we will characterize the splitting functions in terms of certain functions $f(z_{12}, \bar{z}_{12}, \sigma, t)$, and the corresponding OPE coefficients are then determined by integrals of these functions against the measure $\rmd \mathcal{T} = \rmd t\, t^{\D_1-1}\,\left(1-t\right)^{\D_2-1}$. Examples are worked out explicitly in the next section.

Using the parameterization~(\ref{eq:collinearlimitpar1}) for the collinear limit $1||2$, with
\begin{align}
\label{collinearpar2}
     P = \frac{\Lambda\omega}{2} \left(1+z_1 \zb_1, z_1+\zb_1, -i\left(z_1-\zb_1\right), 1-z_1 \zb_1\right)
\end{align}
and taking $z_{12}, \zb_{12} \rightarrow 0$, it follows from the singular behavior of the amplitudes in~(\ref{eq:collineartree}) that the corresponding celestial amplitude behaves as
\begin{align}
    &\tilde{\mathcal{A}}_n^{(0)}\left(\D_1,\sigma_1;\D_2,\sigma_2; \dots; \D_n,\sigma_n\right)\\ &\qquad\qquad\xrightarrow[]{z_{12},\zb_{12}\to 0} \int_{0}^{1} \rmd\mathcal{T} \int_0^\infty \rmd\omega \, \omega^{\D_1+\D_2-1} \sum_{\sigma = \pm }\,\Big[\text{Split}^{(0)}_{-\sigma}(t;1^{\sigma_1},2^{\sigma_2})\nonumber\\
    \nonumber&\hspace{6cm}\times\prod_{j\neq 1,2}\int_0^\infty \rmd\omega_j \,\omega_j^{\D_j-1} \mathcal{A}_{n-1}^{(0)} \left(P,\sigma;p_3,\sigma_3; \dots; p_n,\sigma_n\right)\Big]
\end{align}
where we use the measure defined above and we have suppressed all color dependence in the above equation because it is identical to~(\ref{eq:collineartree}). Noting that all of the tree-level splitting functions (\ref{eq:collineartree}) take the form
\begin{align}
\label{eq:treegenericsplit}
    \text{Split}^{(0)}_{-\sigma}(t;1^{\sigma_1},2^{\sigma_2}) = \frac{1}{\omega} f^{(0)}\left(z_{12},\zb_{12},\sigma,t\right)
\end{align}
(this equation is meant to define $f^{(0)}$, whose dependence on $\sigma_1, \sigma_2$ we suppress) we see that
\begin{multline}
    \label{eq:collinearlimitcelestial}
     \tilde{\mathcal{A}}_n^{(0)}\left(\D_1,\sigma_1;\D_2,\sigma_2;\dots; \D_n,\sigma_n\right)  \\
     \xrightarrow[]{z_{12},\zb_{12}\to 0}\sum_{\sigma = \pm}\tilde{\mathcal{A}}_{n-1}^{(0)}\left(\D_1+\D_2-1,\sigma;\D_3,\sigma_3;\dots; \D_n,\sigma_n\right)\int_{0}^{1} \rmd\mathcal{T}\,f^{(0)}\left(z_{12},\zb_{12},\sigma,t\right).
\end{multline}
This behavior of celestial amplitudes implies the same for the corresponding CCFT correlation functions:
\begin{align}
    \label{eq:OPEcelestial}
     &\an{\mO_{\D_1,\sigma_1}\left(z_1,\zb_1\right)\mO_{\D_2,\sigma_2}\left(z_2,\zb_2\right) \dots \mO_{\D_n,\sigma_n}\left(z_n,\zb_n\right)}^{(0)}\xrightarrow[]{z_{12},\zb_{12}\to 0}\\
     &\qquad\sum_{\sigma = \pm}\an{\mO_{\D_1+\D_2-1,\sigma}\left(z_1,\zb_1\right) \mO_{\D_3,\sigma_3}\left(z_3,\zb_3\right)\dots \mO_{\D_n,\sigma_n}\left(z_n,\zb_n\right)}^{(0)}
       \int_{0}^{1}   \rmd\mathcal{T}\,f^{(0)}\left(z_{12},\zb_{12},\sigma,t\right).
\nonumber
\end{align}
The tree-level OPE coefficients were determined in this manner in~\cite{Pate:2019lpp,Fan:2019emx} for gluons and gravitons.

This formalism readily extends to loop level where the one-loop splitting functions have the generic form (cf.~(\ref{plusplusgoestoplus}))
\begin{align}
\label{eq:loopgenericsplit}
    \alpha \, \text{Split}^{(1)}_{-\sigma}(t,\omega;1^{\sigma_1},2^{\sigma_2}) = \frac{1}{\omega} \sum_{r=0}^2 a f_r^{(1)}\left(z_{12},\zb_{12},\sigma, t\right) \ln^r \left( -\omega^2 t^2(1-t)^2z_{12}\zb_{12}\frac{\Lambda^2}{\mu^2}\right)
\end{align}
for certain functions $f_r^{(1)}$ (see section~\ref{sect:oneloopOPE} for specific examples). The combination $\omega^2 t^2(1-t)^2z_{12}\zb_{12}$ that occurs in the argument of the logarithms is Lorentz invariant in (1,3) signature spacetime. We can rewrite this in a convenient way by noting that
\begin{multline}
    \label{eq:Dijmotivate}
    \int_0^\infty d\omega \, \omega^{\D_1+\D_2-1}\, t^{\D_1-1}\left(1-t\right)^{\D_2-1} \ln \left(-\omega^2 t^2(1-t)^2z_{12}\zb_{12}\frac{\Lambda^2}{\mu^2}\right) (\dots)\\
     :=  2\mathcal{\hat{D}}_{12}\int_0^\infty \rmd\omega \, \omega^{\D}\, t^{\D_1-1}\left(1-t\right)^{\D_2-1}  (\dots) \Big|_{\D=\D_1+\D_2-1}
\end{multline}
where $(\dots)$ is an arbitrary function of $\omega$ and $t$ and in the second line we have defined a new ``covariant derivative'' operator\footnote{This operator transforms appropriately only in the OPE limit $z_{ij},\zb_{ij}\rightarrow 0$.}${}^{,}$
\begin{equation}
\label{eq:covariantderivative}
    \mathcal{\hat{D}}_{ij}=\partial_{\Delta_i}+\partial_{\Delta_j}+\partial_{\Delta}+\frac{1}{2}\ln\left(- z_{ij}\zb_{ij}\frac{\Lambda^2}{\mu^2}\right).
\end{equation}
Importantly, $\D$ must be assumed to be independent of $\D_1, \D_2$ before taking the derivative. Also, it is to be understood that we always take $\Delta = \Delta_1 + \Delta_2 - 1$ after differentiating. Using~(\ref{eq:collinearloop}), we can express the collinear behavior of the loop-level celestial amplitude as
\begin{align}
\nonumber
    &\tilde{\mA}_n^{(0)}(\D_1,\sigma_1;\D_2,\sigma_2; \dots \D_n,\sigma_n)+ \tilde{\mA}_n^{(1)}(\D_1,\sigma_1;\D_2,\sigma_2; \dots \D_n,\sigma_n) \xrightarrow[]{z_{12},\zb_{12}\to 0}  (\ref{eq:collinearlimitcelestial})\,+ \\
     &\qquad\qquad\sum_{\sigma = \pm}  \tilde{\mA}_{n-1}^{(1)}\left(\D_1+\D_2-1,\sigma;\D_3,\sigma_3; \dots \D_n,\sigma_n\right) \int_0^1 \rmd\mathcal{T}\,f^{(0)}\left(z_{12},\zb_{12},\sigma,t\right)\\
     &\nonumber \qquad\qquad +\sum_{r=0,\sigma = \pm}^{r=2}2^r\mathcal{\hat{D}}_{12}^r\tilde{\mA}_{n-1}^{(0)}\left(\D,\sigma;\D_3,\sigma_3; \dots \D_n,\sigma_n\right) \int_0^1 \rmd\mathcal{T} \,f^{(1)}_r\left(z_{12},\zb_{12},\sigma,t\right).
\end{align}
The operator $\partial_{\D}$ in $\mathcal{\hat{D}}_{12}$ acts only on the celestial amplitude while $\partial_{\D_1}$ and $\partial_{\D_2}$ act only on the integral (recall that the measure $\rmd\mathcal{T}$ contains $\D_1$ and $\D_2$). Rewritten in terms of correlation functions, this is
\begin{align}
     &\an{\mO_{\D_1,\sigma_1}\left(z_1,\zb_1\right)\mO_{\D_2,\sigma_2}\left(z_2,\zb_2\right) \dots \mO_{\D_n,\sigma_n}\left(z_n,\zb_n\right)} \xrightarrow[]{z_{12},\zb_{12}\to 0} \\
     &\quad\sum_{\sigma = \pm}\nonumber \int_0^1 \rmd\mathcal{T} \,f^{(0)}\left(z_{12},\zb_{12},\sigma,t\right) \an{\mO_{\D_1+\D_2-1,\sigma}\left(z_1,\zb_1\right)\mO_{\D_3,\sigma_3}\left(z_3,\zb_3\right) \dots \mO_{\D_n,\sigma_n}\left(z_n,\zb_n\right)}\\
     &\nonumber \quad +a \sum_{r,\sigma = \pm}2^r\mathcal{\hat{D}}_{12}^r \int_0^1 \rmd\mathcal{T}\, f^{(1)}_r \left(z_{12},\zb_{12},\sigma,t\right) \an{\mO_{\D,\sigma}\left(z_1,\zb_1\right)\mO_{\D_3,\sigma_3}\left(z_3,\zb_3\right) \dots \mO_{\D_n,\sigma_n}\left(z_n,\zb_n\right)} \\
     &\quad+ \mO\left(a^2\right).\nonumber
\end{align}
We emphasize that in the second line, we've replaced the tree-level correlator by the full correlator since we're only working to $\mO(a)$. This approximation is crucial for us to be able to infer the OPE
\begin{multline}
    \label{eq:genericOPE}
    \mO_{\D_1, \sigma_1}\left(z_1, \zb_1\right) \mO_{\D_2, \sigma_2}\left(z_2, \zb_2\right)
    \sim \sum_{\sigma = \pm} \int_0^1 \rmd\mathcal{T} \,f^{(0)}\left(z_{12},\zb_{12},\sigma,t\right) \mO_{\D_1+\D_2-1,\sigma}\left(z_2,\zb_2\right)\\
     +a\sum_{r,\sigma = \pm}2^r\mathcal{\hat{D}}_{12}^r\int_0^1 \rmd\mathcal{T}\, f^{(1)}_r \left(z_{12},\zb_{12},\sigma,t\right) \mO_{\D,\sigma}(z_2,\bar{z}_2).
\end{multline}

In the computation of all-loop OPEs in section~\ref{sec:higgsregulated}, we will also encounter a splitting function of the form
\begin{multline}
    \label{eq:allloopgenericsplit}
     \text{Split}_{-\sigma}(t,\omega;1^{\sigma_1},2^{\sigma_2}) = \frac{f^{(0)}\left(z_{12},\zb_{12},\sigma,t\right)}{\omega}\\
     \times \exp\left[a \sum_{r=0}^2f_r^{(1)}\left(z_{12},\zb_{12},\sigma, t\right) \ln^r \left(-\omega^2 t^2(1-t)^2z_{12}\zb_{12}\frac{\Lambda^2}{\mu^2}\right)\right],
\end{multline}
which gives the all-loop OPE
\begin{multline}
    \label{eq:genericallloopOPE}
    \mO_{\D_1, \sigma_1}\left(z_1, \zb_1\right) \mO_{\D_2, \sigma_2}\left(z_2, \zb_2\right) \\
    \sim \sum_{\sigma = \pm} \int_0^1 \exp\left[a\sum_{r=0}^2f_r^{(1)}\left(z_{12},\zb_{12},\sigma, t\right) 2^r\dhat_{12}^r\right] f^{(0)}\left(z_{12},\zb_{12},\sigma,t\right) \mO_{\D,\sigma}\left(z_2,\zb_2\right)\rmd \mathcal{T}.
\end{multline}

\subsection{One-loop gluon OPEs}
\label{sect:oneloopOPE}

In this section we will employ the formalism developed above to compute the one-loop correction to the OPE of two positive helicity gluons. The relevant splitting functions expressed in the parameterization of~(\ref{eq:momentaparameterization}), (\ref{eq:anglesquare}) are
\begin{align}
    &\text{Split}^{(0)}_{-}(t;1^{+},2^{+})=\frac{1}{t(1-t)}\frac{1}{\omega} \frac{1}{z_{12}}\,,\\
    &\text{Split}^{(1)}_{-}(t;1^{+},2^{+})= \,\frac{1 }{6}\frac{1}{\omega}\frac{1}{z_{12}}\left(1+\frac{n_s}{N_c}-\frac{n_f}{N_c}\right)+\frac{1}{\omega}\frac{1}{t\left(1-t\right)z_{12}} G^n\,,\\
    &\text{Split}^{(1)}_{+}(t;1^{+},2^{+})= \,\frac{1}{6}\frac{1}{\omega}\frac{\zb_{12}}{z^2_{12}}\left(1+\frac{n_s}{N_c}-\frac{n_f}{N_c}\right),
\end{align}
where
\begin{align}
    \al\, G^n &= \al\,\hat{c}_{\Gamma}\left[-\frac{1}{\e^2}\left(\frac{-1}{t^2(1-t)^2z_{12}\zb_{12}\omega^2 \frac{\Lambda^2}{\mu^2}}\right)^{\epsilon}+2\ln(t)\ln(1-t)-\frac{\pi^2}{6}\right] \\
    &= -\frac{a}{2\e^2}+\frac{a}{2\e}\left[c_E+\ln\left(-t^2(1-t)^2\omega^2z_{12}\zb_{12} \frac{\Lambda^2}{\mu^2}\right)\right]-\frac{a}{4}\ln^2\left(-t^2(1-t)^2\omega^2z_{12}\zb_{12}\frac{\Lambda^2}{\mu^2}\right)\\
    &\quad -\frac{a}{24}\left[\pi^2+6c_E^2-24\ln(t)\ln(1-t)+12c_E\ln\left(-t^2(1-t)^2\omega^2z_{12}\zb_{12}\frac{\Lambda^2}{\mu^2}\right)\right].
\end{align}
We have introduced $c_{E}=\gamma_{E}-\ln(4\pi)$, where $\gamma_{E}$ is the Euler-Mascheroni constant, and we expanded everything to $\mO\left(\e\right)$. From this, we can read off that the functions $f^{(1)}_r$ appearing in~(\ref{eq:loopgenericsplit}) are
\begin{align}
    \label{eq:+++f}
    &f^{(0)} \left(z_{12},\zb_{12},+,t\right) = \frac{1}{z_{12}}\frac{1}{t\left(1-t\right)}\,, \\
    & f^{(1)}_0 \left(z_{12},\zb_{12},+,t\right) =\frac{1}{2} \frac{1}{z_{12}}\left[\frac{1}{3}\left(1+\frac{n_s}{N_c}-\frac{n_f}{N_c}\right)+\frac{1}{t\left(1-t\right)}\left(-\frac{1}{\epsilon^2}+\frac{c_E}{\epsilon}\right.\right.\\
    &\left.\left. \hspace{6.3cm}-\frac{1}{12}\left(\pi^2+6c_E^2-24 \ln t \ln \left(1-t\right)\right)\right)\right],\nonumber\\
    & f^{(1)}_1 \left(z_{12},\zb_{12},+,t\right)  = \frac{1}{2}\frac{1}{t\left(1-t\right)z_{12}}\left[\frac{1}{\epsilon}-c_E\right],\\
     & f^{(1)}_2 \left(z_{12},\zb_{12},+,t\right)  = -\frac{1}{4}\frac{1}{t\left(1-t\right)z_{12}}\,,\\
     & f^{(1)}_0 \left(z_{12},\zb_{12},-,t\right) =\frac{1}{6} \frac{\zb_{12}}{z^2_{12}}\left(1+\frac{n_s}{N_c}-\frac{n_f}{N_c}\right).
\end{align}
According to~(\ref{eq:genericOPE}), the OPE coefficients are obtained by integrating these functions with the measure $\rmd\mathcal{T}$. Using the formulas
\begin{align}
    &\int_0^1 \rmd\mathcal{T} = \frac{\left(\D_1-1\right)\left(\D_2-1\right)\text{B}\left(\D_1{-}1,\D_2{-}1\right)}{\left(\D_1+\D_2-1\right)\left(\D_1+\D_2-2\right)}\,,\\
    &\int_0^1 \rmd\mathcal{T} \frac{1}{t\left(1-t\right)} = \text{B}\left(\D_1{-}1,\D_2{-}1\right), \\
    &\int_0^1 \rmd\mathcal{T} \frac{1}{t\left(1-t\right)}\ln t \ln\left(1-t\right) = \partial_{\D_1}\partial_{\D_2}\text{B}\left(\D_1{-}1,\D_2{-}1\right),
\end{align}
where $\text{B}(a,b) = \Gamma(a)\Gamma(b)/\Gamma(a+b)$, we obtain the OPE
\begin{empheq}[box=\fbox]{multline}
\label{eq:++OPEfinal}
    \mO^a_{\D_1,+}\left(z_1, \zb_1\right) \mO^b_{\D_2,+}\left(z_2,\zb_2\right) \sim \frac{if^{abc}}{z_{12}} \Big[1+\frac{a}{2}\Big( C^{(1)}_{0,+} + C^{(1)}_{1,+} \mathcal{\hat{D}}_{12}\\
 +  C^{(1)}_{2,+}\mathcal{\hat{D}}_{12}^2\Big)\Big]C^{(0)}_{+}\mO^c_{\D,+}\left(z_2,\zb_2\right)
     +if^{abc}\,\frac{a}{2} \,\frac{\zb_{12}}{z^2_{12}} C^{(1)}_{0,-} \,\mO^c_{\D_1+\D_2-1,-}\left(z_2,\zb_2\right)
\end{empheq}
where $C^{(0)}_{+} = \text{B}\left(\D_1-1,\D_2-1\right)$ is the tree-level OPE coefficient~\cite{Pate:2019lpp,Fan:2019emx} and
\begin{align}
    &C^{(1)}_{0,+} = H_{0,+}^{(1)} + S_{0,+}^{(1)}\,, \qquad S_{0,+}^{(1)} =-\frac{1}{\e^2}+\frac{c_E}{\e}-\frac{c_E^2}{2}\,, \nonumber \\
    &H_{0,+}^{(1)} = \frac{1}{3}\left(1+\frac{n_s}{N_c}-\frac{n_f}{N_c}\right)\frac{\left(\D_1-1\right)\left(\D_2-1\right)}{\left(\D_1+\D_2-1\right)\left(\D_1+\D_2-2\right)}-\frac{\pi^2}{12}+2\partial_{\D_1}\partial_{\D_2}\,,\label{eq:++OPEfinalcoeffs}\\
    &C^{(1)}_{1,+} = 2\left(\frac{1}{\e}-c_E\right), \qquad C^{(1)}_{2,+} = -2\,, \qquad C^{(1)}_{0,-} =\frac{1}{3}\Big(1+\frac{n_s}{N_c}-\frac{n_f}{N_c}\Big)\text{B}\left(\D_1,\D_2\right). \nonumber
\end{align}
We pause here to point out a few salient features of this OPE. First, it is interesting to note that it was possible to write the loop corrections as operators acting on the tree-level OPEs. The one exception is the term proportional to $\mO_{\D_1+\D_2-1,-}$ in~(\ref{eq:++OPEfinal}). This has no counterpart at tree-level in Yang-Mills theory. Secondly, the OPE now has second order poles in the $\Delta_i$ in contrast to the simple poles at tree-level.

The splitting behavior considered so far was for bare amplitudes. Upon renormalization, the splitting functions are modified as (following the renormalization scheme in section 2 of~\cite{Bern:1995ix})
\begin{equation}
    \text{Split}^{(1)}_{\text{ren.}}(t;a^{\sigma_a},b^{\sigma_b})=\text{Split}^{(1)}(t;a^{\sigma_a},b^{\sigma_b})-\frac{\hat{c}_{\Gamma}}{\epsilon}\Big(\frac{11}{6}-\frac{1}{3}\frac{n_f}{N_c}-\frac{1}{6}\frac{n_s}{N_c}\Big)\text{Split}^{(0)}(t;a^{\sigma_a},b^{\sigma_b})\,.\label{renorm}
\end{equation}
This modifies the OPE to
\begin{align}
    \label{eq:renormalizedOPE}
    \mO^a_{\D_1,+}\left(z_1, \zb_1\right) \mO^b_{\D_2,+}\left(z_2,\zb_2\right) \sim (\ref{eq:++OPEfinal}) + \frac{if^{abc}}{z_{12}}\frac{a}{4}C_{1,+}^{(1)} C_{+}^{(0)}\Big(\frac{11}{6}-\frac{1}{3}\frac{n_f}{N_c}-\frac{1}{6}\frac{n_s}{N_c}\Big)\mO_{\D_1+\D_2-1,+}^{c}\,.
\end{align}

\subsection{Hard/soft factorization and the hard OPE}
\label{sec:hardsoftfact}

The OPE derived in section~\ref{sect:oneloopOPE} is IR divergent. A natural way to extract a finite OPE is by appealing to the hard/soft factorization theorem for massless gauge theories~\cite{Sen:1982bt,Dixon:2008gr,Gardi:2009qi,Gardi:2009zv,Becher:2009qa,Becher:2009cu,Feige:2014wja,Ma:2019hjq} which takes the form
\begin{align}
    &\bigg[\mathcal{A}^{\text{ren}}_{n}\Big(\frac{p_i}{\mu},\alpha_s(\mu)\Big)\bigg]^{a_1a_2...a_n}=\bigg[\textbf{Z}_n\Big(\frac{p_i}{\mu},\alpha_s(\mu)\Big)\bigg]^{a_1a_2...a_n}_{a_1'a_2'...a_n'}\bigg[\mathcal{H}_n\Big(\frac{p_i}{\mu},\alpha_s(\mu)\Big)\bigg]^{a_1'a_2'...a_n'}\,.\label{eq:hardsoftmomentum}
\end{align}
Here $\mA^{\text{ren}}_{n}$ is the color-dressed, UV renormalized, $n$-gluon amplitude in dimensional regularization and $a_i$ are color indices. All IR divergences are contained in the color operator $\b{Z}_n$. The hard operators are defined by
\begin{align}
\label{eq:hardamplitudedef}
\braket{H^{a_1}_{\Delta_1,\sigma_1}\left(z_1, \zb_1\right) \dots H^{a_n}_{\Delta_n,\sigma_n}\left(z_n,\zb_n\right)} := \int \prod_{i=1}^n d\omega_i\, \omega_i^{\D_i-1} \mathcal{H}_n^{a_1\dots a_n}\,.
\end{align}
Note that $\b{Z}_n = 1 + \mO\left(a\right)$ which implies that
\begin{align}
\label{eq:treelevelhardops}
    \braket{H^{a_1}_{\Delta_1,\sigma_1}\left(z_1, \zb_1\right) \dots H^{a_n}_{\Delta_n,\sigma_n}\left(z_n,\zb_n\right)}^{(0)} = \tilde{\mA}_n^{(0)}\,.
\end{align}
This lets us infer the OPE
\begin{multline}
    H^{a}_{\Delta_1,+}(z_1,\zb_1)H^{b}_{\Delta_2,+}(z_2,\zb_2)\sim \frac{if^{abc}}{z_{12}}\left[\text{B}(\Delta_1-1,\Delta_2-1)H^{c}_{\Delta_1+\Delta_2-1,+}(z_2,\zb_2)\right.\\
    \left.\quad +a\,C_H(z_{12}, \zb_{12}, \D_1, \D_2,\D,\sigma)\, H^{c}_{\Delta,\sigma}(z_2,\zb_2)\right]+\mathcal{O}(a^2)\,,\label{eq:hardOPE}
\end{multline}
where $C_H(z_{12}, \zb_{12}, \D_1, \D_2,\D,\sigma)$ is the one-loop correction that we must determine\footnote{For simplicity we have already presumed the color structure of $C_H$ to be $f^{abc}$, which a more general analysis demonstrates to be correct.}. We will typically suppress the arguments of $C_H$ in the rest of the paper. In the parameterization~(\ref{eq:momentaparameterization}), the universal soft factor can be written as\footnote{We will work in the dipole approximation in this paper since deviations are expect only start at three loops and beyond.}
\begin{align}
    \label{eq:softfactorfactorization}
    \textbf{Z}_n\Big(\frac{p_i}{\mu},\alpha(\mu)\Big) =e^{n\, J} \textbf{Z}_n\Big(\frac{\Lambda^2}{\mu^2}\left|z_{ij}\right|^2,\alpha(\mu)\Big) \prod_{k=1}^n \left(\frac{\omega_k}{\mu}\right)^{-\frac{K}{2}}
\end{align}
where
\begin{align}
\label{eq:Jdef}
   K&=\frac{1}{2}\int_0^{\mu^2}\frac{d\lambda^2}{\lambda^2}\gamma\left(\alpha(\lambda^2)\right) = -a\left(\frac{1}{\epsilon}-c_{E}\right) + \mO\left(a^2\right) =-\frac{a}{2} C_{1,+}^{(1)}
\end{align}
and
\begin{align}
\label{eq:Kdef}
   J&=\frac{1}{8}\int_0^{\mu^2}\frac{d \lambda^2}{\lambda^2}\ln\Big(\frac{\lambda^2}{\mu^2}\Big)\gamma(\alpha)-\frac{1}{2}\gamma_{J}(\alpha)\\
   \nonumber&= \frac{a}{2}\left(-\frac{1}{\e^2}+\frac{c_E}{\epsilon}-\frac{c_E^2}{2}\right)+\frac{a}{2}\left(-\frac{1}{\epsilon}+c_E\right)\left(\frac{11}{6}-\frac{1}{3}\frac{n_f}{N_c}-\frac{1}{6}\frac{n_s}{N_c}\right)\\
&=\frac{a}{2}S_{0,+}^{(1)}+\frac{a}{4}\left(\frac{11}{6}-\frac{1}{3}\frac{n_f}{N_c}-\frac{1}{6}\frac{n_s}{N_c}\right)C_{1,+}^{(1)}\,.\nonumber
\end{align}
In arriving at the above results, we have used the one-loop values of the cusp anomalous dimension $\gamma\left(\alpha\right)$ and collinear anomalous dimension of gluons $\gamma_{J_i}$~\cite{DelDuca:2014cya} appropriate for this theory, performed the integral after using the one-loop RG equation to compute the running coupling. Finally, we have expressed them in terms of the OPE coefficients defined in~(\ref{eq:++OPEfinalcoeffs}). The soft factor can be written as the correlator of $2$D vertex operators. Building upon observations made in~\cite{Arkani-Hamed:2020gyp,Kalyanapuram:2020epb,Nande:2017dba}, it was shown in~\cite{Magnea:2021fvy, Nastase:2021izh,Gonzalez:2021dxw} that the soft factor in non-abelian gauge theories can be expressed as a correlator of vertex operators, i.e.
\begin{align}
    e^{n\,J}\bigg[\textbf{Z}_n\Big(\frac{p_i}{\mu},\alpha(\mu)\Big)\bigg]^{a_1a_2...a_n}_{a_1'a_2'...a_n'} = \Big\langle[\b{V}_{\kappa_1}(z_1,\bar{z}_1)]^{a_1}_{a_{1}'}...[\b{V}_{\kappa_n}(z_n,\bar{z}_n)]^{a_n}_{a_{n}'}\Big\rangle\,.
\end{align}
Consequently, the celestial amplitude factorizes as
\begin{align}
    &\tilde{\mathcal{A}}^{a_1...a_n}=\Big\langle[\b{V}_{\kappa_1}(z_1,\bar{z}_1)]^{a_1}_{a_{1}'}...[\b{V}_{\kappa_n}(z_n,\bar{z}_n)]^{a_n}_{a_{n}'}\Big\rangle\Big\langle H^{a_1'}_{\sigma_1,\Delta_1'}(z_1,\bar{z}_1)...H^{a_n'}_{\sigma_n,\Delta_n'}(z_n,\bar{z}_n)\Big\rangle\,.\label{celestialfact}
\end{align}
Note that the conformal dimensions of the hard operators are $\D_i' = \D_i -K$. This shift is due to the factor $\prod_{i=1}^n \left(\frac{\omega_i}{\mu}\right)^{-K}$ in~(\ref{eq:softfactorfactorization}). The OPE of the vertex operators was derived in~\cite{Magnea:2021fvy}\footnote{Our vertex operators differ by a factor of $e^{J_i}$ from those in~\cite{Magnea:2021fvy}.}. For the purposes of this paper, it is convenient to express the OPE in the form
\begin{align}
\label{eq:vertexOPE1}
     f^{a'b'c'}[\b{V}_{\kappa_1}(z_1,\bar{z}_1)]^{a}_{a'}\,\, [\b{V}_{\kappa_2}(z_2,\bar{z}_2)]^{b}_{b'}\, \sim \, e^{J}\left(z_{12}\zb_{12}\right)^{-\frac{K}{2}}f^{abc}[\b{V}_{\kappa}(z_1,\bar{z}_1)]_{c'}^{c}\,.
\end{align}
This form of the OPE can be derived from the OPE for $[\b{V}_{\kappa_1}(z_1,\bar{z}_1)]^{a}_{a'}\,\, [\b{V}_{\kappa_2}(z_2,\bar{z}_2)]^{b}_{b'}$ given in~\cite{Magnea:2021fvy} by contracting it with $f^{a'b'c'}$ and using the Jacobi identity. Expanding to first order in $a$, we arrive at
\begin{align}
\label{eq:vertexOPE2}
      &f^{a'b'c'}[\b{V}_{\kappa_1}(z_1,\bar{z}_1)]^{a}_{a'}\,\, [\b{V}_{\kappa_2}(z_2,\bar{z}_2)]^{b}_{b'} \sim \\
      \nonumber&\qquad f^{abc} \left(1+\frac{a}{2}S_{0,+}^{(1)}+\frac{a}{4}\Bigg(\frac{11}{6}-\frac{1}{3}\frac{n_f}{N_c}-\frac{1}{6}\frac{n_s}{N_c}\Bigg)C_{1,+}^{(1)}+\frac{a}{4}C_{1,+}^{(1)}\ln \left(z_{12}\zb_{12} \frac{\Lambda^2}{\mu^2}\right)\right)[\b{V}_{\kappa}(z_1,\bar{z}_1)]_{c'}^{c}\,.\nonumber
\end{align}
Introducing the hard/soft decomposition of the operators
\begin{align}
      \mathcal{O}^{a}_{\Delta,\sigma}(z,\bar{z})=[\b{V}_{\kappa}(z,\bar{z})]^{a}_{a'}H^{a'}_{\Delta',\sigma}(z,\bar{z})\,,\label{decompoperator}
\end{align}
we can use~(\ref{eq:hardOPE}), (\ref{eq:vertexOPE2}) to express the OPE of two gluons as
\begin{align}
\nonumber
    &\mathcal{O}^{a}_{\Delta_1,+}(z_1,\zb_1)\mathcal{O}^{b}_{\Delta_2,+}(z_2,\zb_2)
    \sim \frac{if^{abc}}{z_{12}}\Bigg[C_+^{(0)} \mO^c_{\Delta_1+\Delta_2-1,+}+\frac{a}{2}\left(S_{0,+}^{(1)}+C_{1,+}^{(1)}\dhat_{12}\right)C_+^{(0)} \mO^c_{\Delta_1+\Delta_2-1,+}\\
    &\qquad\qquad\qquad\qquad+\frac{a}{4}\left(\frac{11}{6}-\frac{1}{3}\frac{n_f}{N_c}-\frac{1}{6}\frac{n_s}{N_c}\right)C_{1,+}^{(1)}C_+^{(0)} \mO^c_{\Delta_1+\Delta_2-1,+}+a\,C_H \,\mO^c_{\Delta,\sigma}\Bigg].
\end{align}
Comparing this with~(\ref{eq:++OPEfinal}), we infer the value of $C_H$ and deduce the hard OPE to be
\begin{empheq}[box=\fbox]{multline}
\label{eq:hardOPEfinal}
    H^{a}_{\Delta_1,+}(z_1,\zb_1)H^{b}_{\Delta_2,+}(z_2,\zb_2)\sim\frac{if^{abc}}{z_{12}}\left[1
      +aH_{0,+}^{(1)}+aC_{2,+}^{(1)}\dhat_{12}^2\right]C_{+}^{(0)}H_{\Delta_1+\Delta_2-1,+}^{c}\\
      +if^{abc}\frac{a}{2}\frac{\zb_{12}}{z^2_{12}}C_{0,-}^{(1)}H_{\D_1+\D_2-1,-}^c\,.
\end{empheq}
Similarly, for $g^{+}g^{-}\rightarrow g^{+}$ the hard OPE reads
\begin{multline}
                  H^{a}_{\Delta_1,+}(z_1,\bar{z}_1)H^{b}_{\Delta_2,-}(z_2,\bar{z}_2)\sim \\
                  \frac{if^{abc}}{\zb_{12}}\left[1-a\left(\frac{\pi^2}{24}-\partial_{\Delta_1}\partial_{\Delta_2}
    +\mathcal{\hat{D}}^2_{12}\right)\right]\text{B}(\Delta_1{+}1,\Delta_2{-}1)H^{c}_{\Delta_1+\Delta_2-1,+}\,,
\end{multline}
and for $g^{+}g^{-}\rightarrow g^{-}$ the hard OPE reads
\begin{multline}
                  H^{a}_{\Delta_1,+}(z_1,\bar{z}_1)H^{b}_{\Delta_2,-}(z_2,\bar{z}_2) \sim \\
                  \frac{if^{abc}}{z_{12}}\left[1-a\left(\frac{\pi^2}{24}-\partial_{\Delta_1}\partial_{\Delta_2}
    +\mathcal{\hat{D}}^2_{12}\right)\right]\text{B}(\Delta_1{-}1,\Delta_2{+}1)H^{c}_{\Delta_1+\Delta_2-1,-}\,.
\end{multline}
In the case of planar $\mathcal{N}=4$ SYM theory one simply omits from~(\ref{eq:hardOPEfinal}) the non-factorizing term to obtain
\begin{multline}
        H^{a}_{\Delta_1,+}(z_1,\zb_1)H^{b}_{\Delta_2,+}(z_2,\zb_2)\sim\\
       \frac{if^{abc}}{z_{12}}\Bigg[1-a\Bigg(\frac{\pi^2}{24}-\partial_{\Delta_1}\partial_{\Delta_2}
    +\hat{\mathcal{D}}^2_{12}\Bigg)\Bigg]\text{B}(\Delta_1{-}1,\Delta_2{-}1)H^{c}_{\Delta_1+\Delta_2-1,+}\,.\label{HardplanarOPE}
\end{multline}

\section{\texorpdfstring{All-loop OPE in planar $\mathcal{N}=4$ SYM}{All loop OPE in planar N=4 SYM}}
\label{sec:allloopOPEs}

\subsection{\texorpdfstring{Amplitudes in Higgs-regulated planar $\mathcal{N}=4$ SYM}{Amplitudes in Higgs-regulated planar N=4 SYM}}
\label{sec:higgsregulated}

In planar $\mathcal{N}=4$ super Yang-Mills (SYM) theory, the exponential structure of IR divergences is captured to all orders by the BDS ansatz~\cite{Bern:2005iz}
  \begin{equation}
    \begin{split}
            M_{n}^{\text{BDS}}(\epsilon) = \exp\left[\sum_{L=1}^{\infty}\al^L\left(\frac{1}{4}\gamma^{(L)}+\frac{\e L}{2}\mathcal{G}^{(L)}\right)M_n^{(1)}\left(L\e\right) +R_n\left(a,s_{ij}\right)+\mathcal{O}\left(\e\right)\right],\label{eq:BDS}
    \end{split}
\end{equation}
where $M_n^{\text{BDS}}(\epsilon) = \frac{\mA_n(\epsilon)}{\mA_n^{(0)}}$ (cf.~(\ref{eq:loopexpansion})), $s_{ij} = \left(p_i+p_j\right)^2$, $\gamma^{(L)}\left(a\right)$ is the $L$-loop cusp anomalous dimension~\cite{Korchemskaya:1992je} and $\mathcal{G}^{(L)}$ is the collinear anomalous dimension~\cite{Magnea:1990zb}. $R_n$ is the remainder function which is nonvanishing only for $n>5$~\cite{Drummond:2007cf, Drummond:2007au, Bern:2008ap, Cachazo:2008hp}. The consistency of this expression with the hard/soft factorization of section~\ref{sec:hardsoftfact} can be checked by noting that the IR divergent part of the one-loop amplitude is
\begin{align}
    \label{eq:IRdiv-loop}
    M_n^{(1)} = -\frac{1}{2\e^2}\sum_{i=1}^n\left(-\frac{\mu^2}{s_{i,i+1}}\right)^{\e}+\mO\left(\e^{0}\right).
\end{align}
The all-loop celestial amplitude at four points has been computed in~\cite{Gonzalez:2020tpi} via the BDS Ansatz. The avatar of the hard/soft factorization at the level of celestial amplitude has been analyzed in~\cite{Gonzalez:2021dxw}.

One feature of the BDS ansatz is that, due to cross terms present in the power series expansion of the exponential, the argument inside the exponential must be computed to increasingly higher orders in $\e$ at higher loop order in order to determine $\mathcal{O}(\epsilon^0)$ contributions to $M_n^{\rm BDS}$. Dimensional regularization also breaks the dual superconformal symmetry of the amplitude.  A way to circumvent both of these difficulties was developed in~\cite{Alday:2009zm} where the authors proposed a Higgs mechanism based regulator, which preserves an appropriately generalized version of dual conformal symmetry. Spontaneously breaking $U(N+M) \longrightarrow U(N)\times U(M)$ in $\mathcal{N}=4$ SYM makes some of the gauge bosons massive. In the limit $N \gg M \gg 1$, the dominant contribution to the scattering of massless $U(M)$ gauge bosons comes from planar graphs with a single external loop of massive gauge bosons, and only massless $U(N)$ bosons propagating inside the loop. Based on their computation at one and two loops and taking inspiration from the BDS ansatz, the authors postulated the following formula for the Higgs-regulated four-point amplitude:
\begin{align}
     \ln M_4(s,t) = &\frac{\gamma\left(a\right)}{4}M_4^{(1)}-\Tilde{\mathcal{G}}(a)\left[\ln\left(\frac{s}{m^2}\right)+\ln\left(\frac{t}{m^2}\right)\right]+\Tilde{c}_4(a)+\mathcal{O}(m^2)\,,\label{eq:bdshiggs4}
\end{align}
where
\begin{equation}
\label{eq:higgs4pt1loop}
    M^{(1)}_4(p_1,p_2,p_3,p_4) = -\frac{1}{2}\left[\ln^2\left(\frac{s_{12}}{m^2}\right)+\ln^2\left(\frac{s_{13}}{m^2}\right)-\ln^2\left(\frac{s_{12}}{s_{13}}\right)\right]+\frac{\pi^2}{2}+\mO(m^2)\,,
\end{equation}
$\gamma\left(a\right) = \sum_{L=0}^{\infty} \gamma^{(L)}\left(a\right)$ while $\tilde{\mathcal{G}}$ is the analog of the collinear anomalous dimension appearing in~(\ref{eq:BDS}), $\Tilde{c}_4(a)$ is a finite additive constant and the coupling in this theory is just the t`Hooft coupling $a$. Further support for this conjecture was provided by three- and four-loop computations performed in~\cite{Henn:2010bk,Henn:2010ir}. Their calculations also supported the expected exponentiation of the five-point amplitude
  \begin{equation}
  \label{eq:bdshiggs5}
    \begin{split}
        \ln{M_5} &= \frac{\gamma(a)}{4}M_5^{(1)}-\frac{\Tilde{\mathcal{G}}_0(a)}{2}\sum_{i=1}^5\ln\left(\frac{s_i}{m^2}\right)+\mO(m^2)
\end{split}
\end{equation}
where $s_i = s_{i,i+1}$ and
\begin{align}
\label{eq:bdshiggs5pt1loop}
    M_5^{(1)} = &-\frac{1}{4}\sum_{i=1}^{5}\left[\ln^2\left(\frac{s_i}{m^2}\right)-2\ln\left(\frac{s_i}{s_{i+1}}\right)\ln\left(\frac{s_{i+2}}{s_{i-2}}\right)\right]+\frac{5\pi^2}{12}+\mathcal{O}(m^2)\,.
\end{align}
is the five-point, one-loop amplitude.

The $m \to 0$ IR divergences of Higgs-regulated amplitudes take the form of powers of $\ln m$, and since $m^a \ln({s}/{m^2})^b \xrightarrow[]{m\to 0}0$ for any positive $a, b$, there are no cross terms present in the power series expansion of the exponential, unlike in dimensional regularization.  At non-zero values of $m$, this provides an example of an all-loop, IR finite amplitude\footnote{Of course, an important order-of-limits issue is hiding here: when integrating over all values of energy $s$, regions of exponentially large and small values $s \sim m^2 e^{\pm \frac{1}{m^a}}$ do contribute to a term like $m^a \ln\left(\frac{s}{m^2}\right)$. Hence it is necessary to take $m\rightarrow 0$ before taking a Mellin transform to the celestial amplitude.}.

It is instructive to compare the Higgs-regulated amplitude to the one we obtain via the dimensionally regularization BDS formula~(\ref{eq:BDS}). Working with $n=4$ and dropping the $\frac{1}{\epsilon}$ and $\frac{1}{\e^2}$ terms we have
\begin{multline}
    \label{eq:4ptBDS}
    \ln M_4^{\text{BDS}}(s,t)|_{\text{finite}} = -\frac{1}{4}\gamma(a)\left[\ln^2\left(-\frac{\mu^2}{s}\right)+\ln^2\left(-\frac{\mu^2}{t}\right)\right]\\
    -\frac{\mathcal{G}(a)}{2}\left[\ln\left(-\frac{\mu^2}{s}\right)+\ln\left(-\frac{\mu^2}{t}\right)\right]+\frac{\gamma(a)}{8}\ln^2\left(\frac{s}{t}\right)+C(a)+\mathcal{O}(\e)\,.
\end{multline}
The amplitude is equal (up to an additive constant and redefinition of collinear and cusp anomalous dimension) to the amplitude in~(\ref{eq:bdshiggs4}) obtained via Higgs regularization.

It is natural to expect that this relationship continues to hold at the level of celestial amplitudes. Testing this expectation requires us to first identify an IR finite celestial amplitude in the dimensionally regulated theory. To this end, we can take advantage of the fact that celestial amplitudes in $\mathcal{N}=4$ SYM factorize as
\begin{equation}
\label{eq:factorization}
   \Tilde{\mathcal{A}}(\{z_i,\bar{z}_i,\Delta_i\}) = \left<\mathcal{V}_{\tau_1}(z_1,\bar{z}_1)\cdots \mathcal{V}_{\tau_n}(z_n,\bar{z}_n)\right>\big<\hat{\mathcal{O}}_{\Delta'_1,l_1}(z_1,\bar{z}_1)\cdots \hat{\mathcal{O}}_{\Delta'_n,l_n}(z_n,\bar{z}_n)\big>\,,
\end{equation}
where $\D_i' = \D_i +\frac{\gamma\left(a\right)}{8\e}$ and $\V_i$ are vertex operators whose dimensions are all equal to $\frac{\gamma\left(a\right)}{8\e}$\footnote{Where $\sigma_i = -\frac{\gamma_1(a)}{8\epsilon}$, $\{\tau_i\}$ are analogs of the $SU(N_c)$ generators $T^a$ given in section~\ref{sec:hardsoftfact}; they may also be seen as a set of vectors in $\mathbb{R}^n$ with the following property
\begin{equation}
    \mathbf{\tau}_i\cdot \mathbf{\tau}_j = \delta_{i,j}-\delta_{i,j+1}-\delta_{i+1,j}+\delta_{i+1,j+1},\hspace{1cm}\sum_i \tau^a_i=0.
\end{equation}
the vertex operators are dressed as
\begin{equation}
    \mathcal{V}_{\mathbf{\tau_{k}}}(z,\bar{z}) = \mathcal{J}(\epsilon,a):\exp{\left[i\mathbb{\tau}_{k}\cdot\Phi(z,\bar{z})\right]}:
\end{equation}
where $\Phi^{a}(z,\bar{z})$ are colored scalar fields.}. This was shown in~\cite{Gonzalez:2021dxw}, where it allowed the authors to define and compute an IR finite celestial amplitude. Mellin transforming~(\ref{eq:bdshiggs4}) we find
\begin{equation}
    \Tilde{\mathcal{A}}_4(\{\Delta_i,\sigma_i,z_i,\Bar{z}_i\}) = \int_0^\infty \prod_{i=1}^4 d\omega_i \, \omega_i^{\D_i-1} M_4 = f(z,\Bar{z})\prod_{i<j}^4z_{ij}^{\frac{h}{3}-h_i-h_j}\Bar{z}_{ij}^{\frac{\Bar{h}}{3}-\Bar{h}_i-\Bar{h}_j}
\end{equation}
where
\begin{multline}
        f(z,\Bar{z}) = 4\left(\frac{m}{\Lambda}\right)^{\beta}g^2\mathcal{C}(a)\delta(z-\Bar{z})\Theta(z-1)z^{\frac{\beta+10}{6}-\Tilde{\mathcal{G}}_0(a)-\frac{i\pi}{4}\gamma(a)}(z-1)^{\frac{\beta+4}{6}}\\
         \times\sqrt{\frac{\pi}{\gamma(a)}}\exp{\left[\frac{1}{\gamma(a)}\left(\frac{\beta}{2}-\frac{\gamma(a)}{4}\ln (-z)-2\Tilde{\mathcal{G}}_0(a)\right)^2+\mathcal{O}(m^2)\right]}\label{higgscel},
\end{multline}
and
\begin{align}
    \beta = \sum_{i=1}^4\Delta_i-4\,, \quad z = \frac{z_{12}z_{34}}{z_{23}z_{41}} \quad \text{and } \quad \mathcal{C}(a) &= \exp{\left(\frac{\pi^2}{8}\gamma(a)+i\pi\Tilde{\mathcal{G}}_0(a)+\Tilde{c}_4(a)\right).}
\end{align}
We refer the reader to appendix~\ref{app:higgreg4pt} for the complete computation. This is in agreement with the four-point correlator of the hard operators found in (2.35) of~\cite{Gonzalez:2021dxw}\footnote{We are only comparing the dependence on kinematics and ignore overall numerical factors and a redefinition of the collinear anomalous dimension.}. We will demonstrate below that this correspondence continues to hold at the level of OPEs. The upshot is that the Higgs-regulated theory seems to offer an efficient way to perform computations involving the hard operators.

Since $z$ is always real and positive due the constraints coming from delta and theta functions in~(\ref{higgscel}), the amplitude has a nontrivial imaginary part arising from $\ln(-z)$. Recently the authors of~\cite{Garcia-Sepulveda:2022lga} discussed unitarity and factorization on the celestial sphere in the context of the three-dimensional $O(N)$ sigma model. They concluded that the imaginary part of the celestial amplitude can be expressed as a sum over residues of the bulk S-matrix at appropriate singularities in $\omega$. It might be interesting to study~(\ref{higgscel}) along those lines.

\subsection{\texorpdfstring{All-loop OPE in the Higgs-regulated $\mathcal{N}=4$ SYM}{All-loop OPE in the Higgs-regulated N=4 SYM}}

In order to compute the OPE we first compute the corresponding splitting function for this theory. Starting with~(\ref{eq:bdshiggs5pt1loop}), parameterizing $p_1$ and $p_2$ as in~(\ref{eq:collinearlimitpar1}), we find\footnote{The subscript $H$ stands for Higgs-regulated and not hard as in section~\ref{sec:hardsoftfact}.}
\begin{equation}
\label{eq:5ptsplit}
     M^{(1)}_5\overset{1\parallel 2}{\longrightarrow}  M^{(1)}_4(P,p_3,p_4,p_5)+r_H^{(1)}(s,m^2;t)+\mO(m^2)
\end{equation}
where
\begin{equation}
\label{eq:higgssplit1loop}
   r_H^{(1)}(s,m^2;t)= -\frac{1}{4}\ln^2\left(\frac{m^2}{\Lambda^2 st(1-t)}\right)+\ln{t}\ln(1-t)-\frac{\pi^2}{12}\,.
\end{equation}
Recall that $M_n^{(1)} = {\mA^{(1)}_n}/{\mA_n^{(0)}}$ and that\footnote{Here we assume particles 1 and 2 have positive helicity.} (cf.~(\ref{eq:collineartree}))
\begin{align}
\mA_5^{(0)}\left(p_1,+;p_2,+,\dots\right) \overset{1\parallel 2}{\longrightarrow} \text{Split}^{(0)}_{-}\left(t;1^+,2^+\right)\mA_4^{(0)}\left(P,+,\dots\right).
\end{align}
We can rewrite~(\ref{eq:5ptsplit}) as
\begin{align}
    \label{eq:N=4splittingfunctions}
    \mA^{(1)}_5 \overset{1\parallel 2}{\longrightarrow} \text{Split}^{(0)}_{-}\left(t;1^+,2^+\right)\mA_4^{(1)}\left(P,+,\dots\right)+ \text{Split}^{(0)}_{-}\left(t;1^+,2^+\right)r_H^{(1)}(s,m^2;t)\mA_4^{(0)}\left(P,+,\dots\right)
\end{align}
which is in agreement with~(\ref{eq:collinearloop}) and~(\ref{plusplusgoestoplus}). The factorizing part of the splitting function doesn't contribute here. Following the procedure outlined in section~\ref{sec:OPEs} we get
\begin{multline}
        {\mathcal{O}}^{a}_{\Delta_1,+}(z_1,\zb_1){\mathcal{O}}^{b}_{\Delta_2,+}(z_{2},\zb_2)\sim\\
        \frac{if^{abc}}{z_{12}}\left[1-
        a\left(\hat{\mathcal{D}}_{12}^2-\partial_{\Delta_1}\partial_{\Delta_2}+\frac{\pi^2}{12}\right)\right]\text{B}(\Delta_1{-}1,\Delta_2{-}1){\mathcal{O}}^{c}_{\Delta-1,+}(z_{2},\zb_2)\label{higgsoneloopOPE}
\end{multline}
with ${\mathcal{D}}_{12}$ defined as in~(\ref{eq:covariantderivative}) with the replacement  $\mu \to m$. We emphasize once again that we treat $\D$ as independent prior to the action of the differential operator. We can now obtain the OPE for the all-loop amplitude in~(\ref{eq:bdshiggs5}). Inserting the collinear behavior of the one-loop amplitude yields
\begin{equation}
    \begin{split}
        M_5(p_1,+;p_2,+;\dots) \overset{1\parallel 2}{\longrightarrow}  r_H(s,m^2;t)M_4(P,+; \dots)\,,
    \end{split}
\end{equation}
where the all-loop splitting function is
\begin{equation}
\label{eq:higgssplitalllloop}
    r_H(s,m^2;t) = \exp\left[\frac{\gamma(a)}{4}r_H^{(1)}(s,m^2;t)+\frac{\Tilde{\mathcal{G}}_0(a)}{2}\ln\left(\frac{m^2}{\Lambda^2 st(1-t)}\right)\right].
\end{equation}
This implies the all-loop OPE (cf.~(\ref{eq:allloopgenericsplit}), (\ref{eq:genericallloopOPE}))
\begin{empheq}[box=\fbox]{multline}
\label{eq:higgsalllloopope}
            {\mathcal{O}}^{a}_{\Delta_1,+}(z_{1},\zb_1){\mathcal{O}}^{b}_{\Delta_2,+}(z_{2},\zb_2) \sim \frac{if^{abc}}{z_{12}}e^{-\frac{\gamma\left(a\right)\pi^2}{48}}\\
            \times \exp\left[-\frac{\gamma(a)}{4}\left(\Hat{\mathcal{D}}^2_{ij}-\partial_{\Delta_i}\partial_{\Delta_j}\right)-\Tilde{\mathcal{G}}_0(a)\Hat{\mathcal{D}}_{ij}\right] \text{B}(\Delta_1{-}1,\Delta_2{-}1){\mathcal{O}}^{c}_{\Delta,+}(z_{2},\zb_2)\,.
\end{empheq}

\subsection{\texorpdfstring{All-loop OPE in dimensionally regulated $\mathcal{N}=4$ SYM}{All-loop OPE in dimensionally regulated N=4SYM}}
\label{sec:dimregope}

In this section, we will derive the all-loop OPE in the dimensionally regulated theory. Based on the discussion presented towards the end of section~\ref{sec:higgsregulated}, we expect the finite part of this OPE to match the one in~(\ref{eq:higgsalllloopope}). Starting with~(\ref{eq:BDS}), and using the collinear behavior of the one-loop amplitude\footnote{Since the collinear behavior is independent of the number of legs, we can work with $n=5$ where the remainder function vanishes}
\begin{equation}
\label{eq:dimregsplit1loop}
    M^{(1)}_n(\epsilon) \longrightarrow M^{(1)}_{n-1}(\epsilon)\,+r_D^{(1)}(s,\e;t)\,,
\end{equation}
where $r^{(1)}_D(s,\e;t)$ is the analog of $r_H^{(1)}\left(s,m^2;t\right)$ (cf.~(\ref{eq:higgssplit1loop})) and is given by
\begin{equation}
\label{eq:oneloopsplitgimreg}
    \begin{split}
        r^{(1)}_D(L\epsilon;t,s)
        &=\hat{c}_{\Gamma} \left[-\frac{1}{(L\epsilon)^2}\left(\frac{-\mu^2}{\Lambda^2t(1-t)s}\right)^{L\epsilon}+2\ln{t}\ln{(1-t)}-\frac{\pi^2}{6}\right]+\mathcal{O}\left({\epsilon}\right)
    \end{split}
\end{equation}
where $\hat{c}_{\Gamma}$ was defined in~(\ref{eq:chat}). Using this in the full amplitude, we extract the all-loop splitting function
\begin{equation}
\label{eq:dimregallloop}
    r_D(s,\e;t) = \exp{\left[\sum_{L=1}^{\infty}\al^L\left(f^{(L)}(L\epsilon)r^{(1)}_D(L\epsilon;t,s)+\mathcal{O}(\epsilon)\right)\right]}
\end{equation}
where $f^{(L)}(\epsilon) = \gamma^{(L)}/4+\epsilon\,\mathcal{G}^{(L)}/2+\mathcal{O}(\epsilon^2)$. Expanding in $\e$ and dropping terms of $\mO\left(\e\right)$, and using~(\ref{eq:allloopgenericsplit}), (\ref{eq:genericallloopOPE}), we get
\begin{align}
\label{eq:comOPE}
       &O^{a}_{\Delta_1,+}(z_1,\zb_1)O^{b}_{\Delta_2,+}(z_2,\zb_2) \\
       &\sim \frac{if^{abc}}{z_{12}}e^{-\frac{\gamma(\alpha)\pi^2}{96}+\frac{i\pi}{8\epsilon}\gamma_1(\al)}\mathcal{J}(\epsilon,\al)\exp\left[\Hat{\mathcal{S}}_{12}+\frac{\gamma_1\left(\al\right)}{4\e}\Hat{\mathcal{D}}_{12}+\mathcal{O}(\epsilon)\right]\text{B}\left(\Delta_1{-}1,\Delta_2{-}1\right)O^{c}_{\D,+}(z_2,\zb_2)\nonumber
\end{align}
with
\begin{equation}
    \Hat{\mathcal{S}}_{ij}(\epsilon,\alpha;z_{ij}) = -\frac{\gamma(\alpha)}{4}\left(\Hat{\mathcal{D}}^2_{ij}-\partial_{\Delta_i}\partial_{\Delta_j}\right)+\frac{\mathcal{G}_0(\alpha)}{2}\Hat{\mathcal{D}}_{ij}\,,
\end{equation}
and the jet function
\begin{equation}
    \mathcal{J}(\epsilon,\al) = \exp\left(-\frac{i\pi}{8\epsilon}\gamma_1(\al)-\frac{\gamma_2(\al)}{8\epsilon^2}-\frac{\mathcal{G}_1(\al)}{4\epsilon}\right)
\end{equation}
 is a non-dynamical factor depending on $\al$ only via the modified anomalous dimensions as defined in~\cite{Gonzalez:2021dxw}
\begin{align}
    \gamma_{m}(\al) = \sum_{L=1}^{\infty}L^{-m}\alpha^L\gamma^{(L)}\,,\qquad \mathcal{G}_{m}(\al) = \sum_{L=1}^{\infty}L^{-m}\alpha^L\mathcal{G}^{(L)}\,.
\end{align}
With the goal of bringing~(\ref{eq:comOPE}) to a form similar to~(\ref{eq:higgsalllloopope}), we note that we can use
\begin{multline}
   \exp\left(\frac{i\pi}{8\epsilon}\gamma_1(\al)+\frac{\gamma_1\left(\al\right)}{4\e} \dhat_{12}\right)\text{B}\left(\Delta_1{-}1,\Delta_2{-}1\right)O^{c,+}_{\D}(z_2,\zb_2) \\
    =\left(\frac{\mu^2}{\Lambda^2z_{12}\zb_{12}}\right)^{-\frac{\gamma_1\left(a\right)}{8\e}} \text{B}\left(\D_1'{-}1,\D_2'{-}1\right)O_{\D+\frac{\gamma}{4\e}}^{c,+}\left(z_2,\zb_2\right)
\end{multline}
to rewrite~(\ref{eq:comOPE}) as
\begin{multline}
    \label{eq:factorizedcomOPE}
     O^{a}_{\Delta_1,+}(z_1,\zb_1)O^{b}_{\Delta_2,+}(z_2,\zb_2) \sim \frac{if^{abc}}{z_{12}}\left[\mathcal{J}(\epsilon,\al)\left(\frac{\mu^2}{\Lambda^2z_{12}\zb_{12}}\right)^{-\frac{\gamma_1\left(\al\right)}{8\e}}\right]e^{-\frac{\gamma(\al)\pi^2}{96}} \\
       \times\exp\left[\Hat{\mathcal{S}}_{12}+\mathcal{O}(\epsilon)\right]\text{B}\left(\Delta'_1{-}1,\Delta'_2{-}1\right)O^{c,+}_{\D+\frac{\gamma}{4\e}}(z_2,\zb_2)\,.
\end{multline}
Finally, we recall from~(\ref{eq:factorization}) that the celestial amplitude factorizes and we can decompose
\begin{equation}
    \mathcal{O}_{\Delta,+}^{a}(z_i,\bar{z}_i) =\,\,:\!\mathcal{V}_{\sigma_i}(z_i,\bar{z}_i)\!:\,\,:\!\hat{\mathcal{O}}_{\Delta',+}^{a}(z_i,\bar{z}_i)\!:
\end{equation}
and that the OPE of the vertex operators is~\cite{Gonzalez:2021dxw},
\begin{equation}
    \mathcal{V}_{\tau_1}(z_1,\zb_1)\mathcal{V}_{\tau_2}(z_2,\zb_2) = \frac{\mathcal{J}(\epsilon,\al)\mathcal{V}_{\tau_1+\tau_2}(z_2,\zb_2)}{\left(z_{12}\zb_{12}\right)^{-\frac{\gamma_1\left(\al\right)}{8\e}}}\,. \label{softOPE}
\end{equation}
This lets us write
\begin{multline}
    \label{eq:factorizedcomOPE2}
     O^{a}_{\Delta_1,+}(z_1,\zb_1)O^{b}_{\Delta_2,+}(z_2,\zb_2) \sim \frac{if^{abc}}{z_{12}}\left(\mathcal{V}_{\tau_1}(z_1,\zb_1)\mathcal{V}_{\tau_2}(z_2,\zb_2) \right) \left(\frac{\mu^2}{\Lambda^2}\right)^{-\frac{\gamma_1\left(\al\right)}{8\e}}e^{-\frac{\gamma(\al)\pi^2}{96}}\\
       \times\exp\left[\Hat{\mathcal{S}}_{12}+\mathcal{O}(\epsilon)\right]\text{B}\left(\Delta'_1{-}1,\Delta'_2{-}1\right)\hat{O}^{c}_{\D',+}(z_2,\zb_2)\,,
\end{multline}
where we note that $\Delta' = \Delta'_1+\Delta'_2-1$. Stripping off the factor
$\mathcal{V}_{\tau_1}(z_1,\zb_1)\mathcal{V}_{\tau_2}(z_2,\zb_2)$ from the LHS, we see that the OPE of the hard operators coincides with the OPE derived from the Higgs-regulated amplitude.

\section{Discussion}
\label{sec:discussion}

\subsection{Role of logarithmic CFTs and operators in CCFT}

The IR finite OPEs obtained in~(\ref{eq:hardOPEfinal}), (\ref{eq:higgsalllloopope}) after soft factorization contain logarithms of $z$ and operators of the form $\partial_{\D}\mO_{\D,\sigma}$. The most familiar OPE which contains logarithms is the free boson OPE. In that case, it is a consequence of working with a weight zero primary whose conformal properties are not well-defined and may be fixed by working with derivatives or exponentials of the operator. A second instance in which logarithms appear is when one performs perturbation theory in a CFT. Logs then appear as a consequence of anomalous dimensions and operator mixing. Upon diagonalizing the dilation operator and summing the perturbative series, one should then find that the logarithms sum to a power law if the conformal symmetry is preserved. It would be of interest to determine whether the all-loop result~(\ref{eq:higgsalllloopope}) is consistent with this scenario.

If not, it raises the more exotic possibility that celestial CFT may be a type of logarithmic CFT (LCFT) (see for example~\cite{Gurarie:1993xq}). In an LCFT states are organized in logarithmic multiplets of rank $r\geq 1$~\cite{Hogervorst:2016itc}. These are built on the top of $r$ primary operators (of the same conformal dimension) which mix under the action of the dilatation operator. The dilatation operator in an LCFT is non diagonalizable rendering the theory non-unitary. Under a conformal transformation parameterized by $f(z)$, a rank-$r$ primary $O_a(z)$ transforms as
\begin{equation}
    O_a(z) = \left(\frac{\partial f(z)}{\partial z}\right)^{\Delta}\sum_{b=a}^{r}\frac{1}{(b-a)!}\ln^{b-a}\left(\partial_z f(z)\right)O_{b}(f(z))\,.\label{eq4.10}
\end{equation}
The logarithmic partners of a primary are defined by~\cite{Hogervorst:2016itc, Flohr:2001zs}
\begin{equation}
    O_a\equiv \frac{1}{(r-a)!}\partial^{r-a}_{\Delta}O_{\Delta}\,.
\end{equation}
The multiplet is
\begin{equation}
    \hat{O}_{\Delta} = \left(\frac{1}{r!}O^{(r)}_{\Delta},\frac{1}{(r-1)!}O^{(r-1)}_{\Delta},\cdots,O^{'}_{\Delta},O_{\Delta}\right)^{T}.
\end{equation}
For some applications of LCFTs, we refer the reader to~\cite{Hogervorst:2016itc,Flohr:2001zs, Nivesvivat:2020gdj}. The all-loop OPE~(\ref{eq:higgsalllloopope}) seems to suggest that at loop order $L$, the corresponding CCFT is described by a rank-$(2L+1)$ multiplet. It would be interesting to analyze this further and to consider the effect this has on the conformal block decomposition in light of~\cite{Hogervorst:2016itc}. As $L\to \infty$ we have an rank-$\infty$ family of log primaries. Not much is known of infinite rank LCFTs; see~\cite{Rasmussen:2004jc} for a discussion.

\subsection{Associativity}

The presence of logarithms prevents us from performing a straightforward check of OPE associativity along the lines of~\cite{Costello:2022upu,Ren:2022sws}. However, in this section we propose a conjectural identity to test OPE associativity when it contains logarithms. The statement resembles the ``double residue formula'' used in~\cite{Ren:2022sws} to check the associativity of the tree-level OPE. Consider a double-OPE limit of three holomorphic operators, $\mO_1(z_1)$, $\mO_2(z_2)$ and $\mO_3(z_3)$. The expansion of $(\mO_1 \mO_2) \mO_3$ can be brought to the form:
\begin{align}\label{eq:logexpansion}
   (\mO_1 \mO_2) \mO_3
    \sim \frac{1}{z_{12}} \frac{1}{z_{23}} \sum_{r,s} \ln^{r}(z_{12}) \ln^{s}(z_{23}) K(1,2,3;r,s;z_3) \,,
\end{align}
where we suppress the color factors for the moment. We conjecture that if the OPE is associative, the following should be satisfied:
\begin{equation}\label{eq:Kidentity}
    \sum_{\substack{r,s \geq 0 \\ r+s=m}} \left( K(1,2,3;r;s;z_3) + K(2,3,1;r;s;z_3) + K(3,1,2;r;s;z_3) \right) = 0 \,,
\end{equation}
where $K\left(2,3,1;r,;s;z_3\right)$ and $K\left(3,1,2;r;s;z_3\right)$ are obtained by taking the OPE limits in different orders.

In order to check if this identity is satisfied by the one-loop OPE of hard operators~(\ref{HardplanarOPE}), we first note that if we restore the color factors then each $K$ has the same color structure, specifically $K\left(i,j,k;r;s;z_3\right)f^{a_ia_jb}f^{ba_kc}$. The identity~(\ref{eq:Kidentity}) is satisfied if $K\left(i,j,k;r;s;z_3\right)$ is symmetric under cyclic permutations of $i,j,k$ by virtue of the Jacobi identity satisfied by the color factors. Evaluating the $K$ coefficients for planar $\mathcal{N}=4$ SYM~(\ref{HardplanarOPE}) we find\footnote{Here the $K^{(n)}_m(\Delta_i,\Delta_j,\Delta_k)$ stands for the coefficient in front of the $n$th logarithmic partner with logarithmic power $m$.}
\begin{align}
       K^{(2)}_0  &= -2\frac{\Gamma(\Delta_i-1)\Gamma(\Delta_j-1)\Gamma(\Delta_k-1)}{\Gamma\left(\Delta_{i}+\Delta_{j}+\Delta_{k}-3\right)}\,, \\
       K^{(1)}_0 &= -2(\partial_{\Delta_i}+\partial_{\Delta_j}+\partial_{\Delta_k})\frac{\Gamma(\Delta_i-1)\Gamma(\Delta_j-1)\Gamma(\Delta_k-1)}{\Gamma\left(\Delta_{i}+\Delta_{j}+\Delta_{k}-3\right)}\,,\\
        K^{(1)}_1 &=  K^{(2)}_0\,,\hspace{5mm}
        K^{(0)}_1 = K^{(1)}_0\,, \hspace{5mm}
        K^{(0)}_2 = -K^{(2)}_0\,.
\end{align}
Interestingly, all of the terms are fully symmetric in the exchange of conformal dimensions and therefore pass the check~(\ref{eq:Kidentity}), this is not true for the single term
\begin{equation}
    \begin{split}
        K^{(0)}_0
    &= \left(\partial^2_{\Delta_i}+\partial^2_{\Delta_j}+\partial^2_{\Delta_k}\right)\frac{\Gamma(\Delta_i-1)\Gamma(\Delta_j-1)\Gamma(\Delta_k-1)}{\Gamma\left(\Delta_{i}+\Delta_{j}+\Delta_{k}-3\right)}\\
    &\qquad+\text{B}(\Delta_{i}{+}\Delta_{j}{-}2,\Delta_k{-}1)\partial_{\Delta_i}\partial_{\Delta_j}\text{B}(\Delta_i{-}1,\Delta_j{-}1)\\
    &\qquad+\text{B}(\Delta_{i}{-}1,\Delta_k{-}1)\partial_{\Delta_i}\partial_{\Delta_k}\text{B}(\Delta_{i}{+}\Delta_{j}{-}2,\Delta_k{-}1)
    \end{split}
\end{equation}
where the exchange symmetry is ruined by the latter double derivative terms of the beta function. In fact one sees that the only two parts that lead to a failure of associativity are the one above and the $F^3$ self-dual term (already considered in~\cite{Costello:2022upu}) regardless of the matter content $\{N_c,n_f,n_s\}$.

\acknowledgments

We are grateful to Shamik Banerjee, Shounak De, Yangrui Hu, Andrew McLeod and Francisco Rojas for useful comments and discussions. This work was supported in part by the US Department of Energy under contract {DE}-{SC}0010010 (Task F) and by Simons Investigator Award \#376208.

\appendix

\section{Calculation of the celestial four-point Higgs-regulated amplitude}
\label{app:higgreg4pt}

For the Higgs-regulated version of $\mathcal{N}=4$ SYM the general MHV (maximum helicity violating) four-point gluon amplitude is simply the tree-level Parke-Taylor amplitude $\mathcal{A}^{(0)}_4$ times the exponential of the function~\cite{Henn:2010ir}
\begin{multline}
     \ln M_4(s,t) = -\frac{1}{8}\gamma(a)\left[\ln^2\left(\frac{s}{m^2}\right)+\ln^2\left(\frac{t}{m^2}\right)\right]-\Tilde{\mathcal{G}}_0(a)\left[\ln\left(\frac{s}{m^2}\right)+\ln\left(\frac{t}{m^2}\right)\right]\\
     +\frac{\gamma(a)}{8}\left[\ln^2\left(\frac{s}{t}\right)+\pi^2\right]+\Tilde{c}_4(a)+\mathcal{O}(m^2)\,.\label{eq1}
\end{multline}
 From~\cite{Nastase:2021izh} we have that any general color-stripped amplitude can be brought into the form
\begin{multline}
         \mathcal{A}(1^{\ell_1},2^{\ell_2},3^{\ell_3},4^{\ell_4}) = z^{\alpha_1}(z-1)^{\alpha_2}\mathcal{B}(s,t)\left(
         \frac{z_{12}}{\bar{z}_{12}}\right)^{-\frac{1}{2}(\ell_1+\ell_2-\ell_3-\ell_4)}\left(
         \frac{z_{13}}{\bar{z}_{13}}\right)^{-\frac{1}{2}(\ell_1+\ell_3-\ell_2+\ell_4)}\\
         \times\left(
         \frac{z_{23}}{\bar{z}_{23}}\right)^{-\frac{1}{2}(\ell_2+\ell_3-\ell_1-\ell_4)}\left(
         \frac{z_{24}}{\bar{z}_{24}}\right)^{-\ell_4}\,.
\end{multline}
The Mellin transform of this formula gives the general four-point celestial gluon amplitude
 \begin{equation}
      \Tilde{\mathcal{A}}_4(\{\Delta_i,\sigma_i,z_i,\Bar{z}_i\}) = f(z,\Bar{z})\prod_{i<j}^4z_{ij}^{\frac{h}{3}-h_i-h_j}\Bar{z}_{ij}^{\frac{\Bar{h}}{3}-\Bar{h}_i-\Bar{h}_j}
 \end{equation}
 where the nontrivial function of conformal cross ratios is given as
 \begin{equation}
     f(z,\Bar{z}) = 2g^2\delta(z-\Bar{z})\Theta(z-1)z^{\frac{\beta+4}{6}+\alpha_1}(z-1)^{\frac{\beta+4}{6}+\alpha_2} \int_0^{\infty}dw\, w^{\frac{\beta-2}{2}}\mathcal{B}(zw,-w)\,.
 \end{equation}
For the MHV four-point gluon amplitude we have $\alpha_1=0$ and $\alpha_2=1$, and we have defined $\beta = \sum_{i=1}^4\Delta_i-4$. For the amplitude~(\ref{eq1}) the kinematic factor $\mathcal{B}(s,t)$ is given as
 \begin{multline}
     \mathcal{B}(s,t) =g^2\left(\frac{s}{m^2}\right)^{-\frac{\gamma(a)}{8}\ln\left(\frac{s}{m^2}\right)}\left(\frac{t}{m^2}\right)^{-\frac{\gamma(a)}{8}\ln\left(\frac{t}{m^2}\right)}\left(\frac{s}{m^2}\right)^{-\Tilde{\mathcal{G}}_0(a)}\\
     \times\left(\frac{t}{m^2}\right)^{-\Tilde{\mathcal{G}}_0(a)}\left(\frac{s}{t}\right)^{\frac{\gamma(a)}{8}\ln\left(\frac{s}{t}\right)}e^{\frac{\pi^2}{8}\gamma(a)+\Tilde{c}_4(a)}\,.
 \end{multline}
Under the change of variables $s=\Lambda z w$, $t=-\Lambda w$ and after some straightforward algebra we get
\begin{multline}
    \mathcal{B}(zw,-w) = \exp{\left(\frac{\pi^2}{8}\gamma(a)+i\pi\Tilde{\mathcal{G}}_0(a)+\Tilde{c}_4(a)\right)}z^{-\Tilde{\mathcal{G}}_0(a)-\frac{i\pi}{4}\gamma(a)}\\
    \times \left(w\frac{\Lambda^2}{m^2}\right)^{-\left(\frac{\gamma(a)}{4}\ln(-z)+2\Tilde{\mathcal{G}}_0(a)+\frac{\gamma(a)}{4}\ln\left(w\frac{\Lambda^2}{m^2}\right)\right)}\,.
\end{multline}
After the substitution $x=w\Lambda^2/m^2$ this then implies that the contribution to the Mellin transform is
\begin{align}
    f(z,\Bar{z}) &\propto \int_0^{\infty}dw\, w^{\frac{\beta-2}{2}}\mathcal{B}(zw,-w)
    =\left(\frac{m}{\Lambda}\right)^{\beta}\int_0^{\infty}dx\,x^{\frac{\beta-2}{2}}\mathcal{B}(z,x)
\end{align}
where
\begin{equation}
   \mathcal{B}(z,x) = \mathcal{C}(a)z^{-\Tilde{\mathcal{G}}_0(a)+\frac{i\pi}{4}\gamma(a)}x^{c_1(a,z)+c_2(a)\ln x}
\end{equation}
for the constants given as
\begin{align}
    \mathcal{C}(a) &= \exp{\left(\frac{\pi^2}{8}\gamma(a)+i\pi\Tilde{\mathcal{G}}_0(a)+\Tilde{c}_4(a)\right)},\label{eq2}\\
    c_1(a,z) &=  -\left(\frac{\gamma(a)}{4}\ln (-z)+2\Tilde{\mathcal{G}}_0(a)\right),\label{eq3}\\
    c_2(a) &= -\frac{\gamma(a)}{4}\,.\label{eq4}
\end{align}
Integrating the above function for Re$(c_2)<0$ (this is always true as the cusp anomalous dimension satisfies $\gamma(a)>0$\footnote{The importance of this fact for the existence of the Mellin transform was emphasized in~\cite{Gonzalez:2021dxw}.}) we get a Gaussian integral which is elementary to evaluate and gives
\begin{equation}
    f(z,\Bar{z}) \propto 2\left(\frac{m}{\Lambda}\right)^{\beta}\mathcal{C}(a)\sqrt{\frac{\pi}{\gamma(a)}}z^{-\Tilde{\mathcal{G}}_0(a)-\frac{i\pi}{4}\gamma(a)}\exp{\left[-\frac{1}{4c_2(a)}\left(\frac{\beta}{2}+c_1(a,z)\right)^2\right]}.
\end{equation}
Substituting in for the constants from~(\ref{eq3}) and~(\ref{eq4}) we have our final result
\begin{multline}
        f(z,\Bar{z}) = 4\left(\frac{m}{\Lambda}\right)^{\beta}g^2\mathcal{C}(a)\delta(z-\Bar{z})\Theta(z-1)z^{\frac{\beta+10}{6}-\Tilde{\mathcal{G}}_0(a)-\frac{i\pi}{4}\gamma(a)}(z-1)^{\frac{\beta+4}{6}}\\
         \times\sqrt{\frac{\pi}{\gamma(a)}}\exp{\left[\frac{1}{\gamma(a)}\left(\frac{\beta}{2}-\frac{\gamma(a)}{4}\ln (-z)-2\Tilde{\mathcal{G}}_0(a)\right)^2+\mathcal{O}(m^2)\right]}\label{eq7.11}
\end{multline}
with the constant $\mathcal{C}(a)$ is given by~(\ref{eq2}).

\bibliography{main7}

\bibliographystyle{JHEP}

\end{document}